\documentclass[10pt,a4paper]{article} % 10pt is ignored!
\usepackage{graphicx}
\usepackage{amsfonts,amsmath,amssymb}
\usepackage{hyperref}
\usepackage{cite}
\newcommand{\bse}{\begin{subequations}}
\newcommand{\ese}{\end{subequations}}

\begin{document}
\title{\bf The DBI critical electric field in stationary non-diagonal geometries}

\date{}
\maketitle
\vspace*{-0.3cm}
\begin{center}
{\bf Leila Shahkarami}\\
\vspace*{0.3cm}
{\it {School of Physics, Damghan University, Damghan 36716-45667, Iran
}} \\
\vspace*{0.3cm}
{\it  {l.shahkarami@du.ac.ir}}
\end{center}

\begin{abstract}
We investigate the critical electric field for the holographic Schwinger effect in general stationary non-diagonal geometries. We show that, in such backgrounds, the DBI reality condition must be formulated in terms of the electric field measured in a local orthonormal frame, since the coordinate components of the field strength need not represent a purely electric field in that frame. The critical field measured in the orthonormal frame takes the universal value \(E_{\mathrm{proper},c}=T_F\), while its coordinate expression depends on the background geometry. The resulting critical field is independently confirmed by the holographic potential analysis, through the disappearance of the potential barrier. We further illustrate the general prescription using the planar limit and the global rotating Myers--Perry--AdS geometry, which exhibit different types of metric mixing.
\end{abstract}
Keywords: Holographic Schwinger effect; Dirac--Born--Infeld action; Critical electric field; Rotating AdS black holes.
%$$$$$$$$$$$$$$$$$$$$$$$$$$$$$$$$$$$$$$$$$$$$$$$$$$$$$$$$$$$$$$$$$$$$$$$$$
\section{Introduction}

The production of particle-antiparticle pairs from the vacuum in the presence of a strong electric field is one of the characteristic nonperturbative phenomena of quantum electrodynamics. In his seminal work, Schwinger showed that a constant electric field produces charged pairs with a rate that is exponentially suppressed at weak fields,
\begin{align}
\Gamma \sim \exp\left(-\frac{\pi m^2}{|qE|}\right),
\end{align}
where $m$ and $q$ denote the mass and charge of the produced particle, respectively \cite{Schwinger}. The exponential suppression reflects the tunneling nature of the process. The strong-field regime, however, lies beyond the range in which the original weak-field expression can be used reliably, and the onset of an unsuppressed instability is more naturally characterized in string theory, where the electric field is bounded by a critical value related to the string tension \cite{string1,string2}.

Gauge/gravity duality \cite{adscft1,adscft2,adscft3,adscft4} provides a useful framework for studying this phenomenon at strong coupling and beyond the weak-field approximation. The holographic Schwinger effect was formulated by Semenoff and Zarembo by considering $\mathcal{N}=4$ supersymmetric Yang--Mills theory on the Coulomb branch and realizing the charged particles as massive $W$ bosons \cite{semenoff}. In the large-$N$ and large-'t Hooft-coupling $\lambda$ limit, the pair-production rate can be obtained holographically and takes a form that exhibits a critical electric field $E_c$,
\begin{align}
\Gamma \sim \exp\left[ -\frac{\sqrt{\lambda}}{2} \left(\sqrt{\frac{E_c}{E}}-\sqrt{\frac{E}{E_c}}\right)^2\right],\qquad E_c=\frac{2\pi m^2}{\sqrt{\lambda}}.
\end{align}
A notable feature of this result is that the critical field coincides with the value obtained from the Dirac--Born--Infeld (DBI) action of the probe D3-brane. This agreement provided one of the first important checks of the holographic description of the Schwinger effect \cite{semenoff}.

Two complementary approaches have subsequently been used to study the holographic Schwinger effect. One is based directly on the DBI action of the flavor or probe brane. For a constant electric field on the brane, the DBI action becomes imaginary once the field exceeds a critical value, and the requirement that the classical action remain real determines $E_c$. The other approach is based on the potential experienced by a quark-antiquark pair. In this description, the total potential contains the rest-energy contribution of the pair, the interaction energy obtained from a fundamental string, and the contribution of the external electric field. Pair production is then interpreted as quantum tunneling through a potential barrier, which decreases as the electric field is increased and disappears at the critical field.

The potential analysis was developed systematically by Sato and Yoshida \cite{potential}. In their initial analysis, evaluating the Coulomb part using a minimal surface extending to the conformal boundary led to a critical field that differed from the DBI result. They showed that this discrepancy originates from the treatment of the string endpoints and that the Coulomb potential should instead be evaluated for a string attached to a probe D3-brane located at a finite radial position. With this prescription, the critical field obtained from the disappearance of the potential barrier agrees exactly with the DBI result, both at zero and finite temperature \cite{potential}. The potential approach was subsequently generalized to a broad class of holographic backgrounds, where the existence of a critical field and its relation to the DBI result were established under general conditions \cite{Sch1}. These developments established the potential analysis and the DBI criterion as two mutually consistent ways of identifying the onset of catastrophic pair production in the standard backgrounds considered so far.

The holographic Schwinger effect has since been investigated in a wide variety of geometries, including confining backgrounds, finite-temperature black holes, anisotropic and magnetized systems, nonconformal models, and rotating backgrounds \cite{Sch2,Sch3,confin1,Sch4,confinrev,Sch5,fadafan,Sch6,dehghani,us2,Sch7,rotSch1,
rotSch2,rotSch3,superus,thermoscaling}. In these applications, the critical electric field is generally obtained from the DBI action and the potential analysis is then used to study the behavior of the potential barrier and the pair-production process. The agreement between the two methods is usually taken as a consistency check of the holographic description.

The success of these two approaches in static and diagonal backgrounds raises a natural question concerning their applicability to more general non-diagonal geometries. In such backgrounds, the induced metric may contain temporal-spatial as well as spatial-spatial off-diagonal components, and the coordinate basis need not coincide with a local orthonormal frame. This distinction is important when introducing an external electric field through the DBI action, since a field that is purely electric in a physical frame may acquire additional components when expressed in the coordinate basis. Rotating and boosted geometries provide particularly relevant examples of this situation. They arise naturally in holographic descriptions of strongly coupled matter with angular momentum or fluid motion and have been extensively studied in the context of heavy-quark observables and the Schwinger effect \cite{rotSch1,rotSch2,rotSch3,superus,thermoscaling,KerrAdSwilson1,MPwilson1,boost1,
boost2,boost3,boost4,global,frameus}. Their physical relevance is also motivated by the substantial angular momentum carried by noncentral heavy-ion collisions and the observed global polarization of emitted hyperons, which provides evidence for vorticity in the produced plasma \cite{star2017,alice,star2023}. These considerations motivate a systematic treatment of the DBI critical field in non-diagonal backgrounds, with particular attention to rotating geometries. A related but distinct line of investigation concerns non-diagonal open string metrics arising from external worldvolume electromagnetic fields. In particular, in \cite{Kim2011} they study such effective geometries and their implications for probe-brane dynamics and transport, rather than the determination of the Schwinger critical field in a non-diagonal bulk spacetime.

Despite the extensive development of holographic Schwinger calculations in nontrivial and rotating backgrounds, the precise determination of the critical electric field from the DBI action in a general stationary non-diagonal geometry has not, to our knowledge, been addressed in a systematic way. In particular, the relation between the DBI reality condition and the locally measured electric field requires some care when the metric contains temporal-spatial cross terms. Consequently, the critical field obtained by directly applying the standard diagonal-background prescription need not be the correct one. 

This motivates us to reconsider the determination of the critical electric field in stationary non-diagonal geometries from a general perspective. Rather than restricting the analysis to a particular black-hole background, we formulate the problem for a general stationary
metric and examine the DBI and total potential descriptions on the same footing. The resulting prescription is then tested in rotating backgrounds, including both planar and global Myers--Perry geometries \cite{MPBH}. This provides a direct consistency check between the two independent descriptions of the onset of the Schwinger instability.

The paper is organized as follows. In Sec.\,2, we first formulate the potential analysis for a general stationary geometry containing temporal-spatial mixing and determine the critical electric field from the disappearance of the potential barrier. We then reconsider the critical field from the DBI action, taking proper account of the non-diagonal metric and formulating the electric field in a local orthonormal frame. The agreement between the two approaches provides an independent check of the resulting critical field. In Sec.\,3, we apply the general analysis to rotating backgrounds, considering both the planar-limit and global Myers--Perry geometries. Finally, Sec.\,4 summarizes our results and discusses their implications for holographic Schwinger processes in stationary non-diagonal geometries.

%$$$$$$$$$$$$$$$$$$$$$$$$$$$$$$$$$$$$$$$$$$$$$$$$$$$$$$$$$$$$$$$$$$$$$$$$$$$$$$$$$$$$$$$$$$
\section{Critical electric field in stationary non-diagonal geometries}

In this section, we investigate the determination of the critical electric field in stationary non-diagonal geometries using two complementary approaches. We first employ the potential analysis to identify the critical field from the disappearance of the potential barrier for a quark-antiquark pair. We then derive the corresponding condition from the DBI action and compare the results obtained by the two methods.

To establish the general prescription, we consider a stationary background whose metric takes the form
\begin{align}
ds^2={}g_{tt}(z)\,dt^2+2g_{t\parallel}(z)\,dt\,dq_\parallel +g_{\parallel\parallel}(z)\,dq_\parallel^2+g_{\perp\perp}(z)\,dq_\perp^2+g_{22}(z)\,dq_2^2 +g_{zz}(z)\,dz^2,
\label{generalmetric}
\end{align}
where $z$ is the holographic radial coordinate with the boundary living at $z=0$, and $q_\parallel$ and $q_\perp$ denote the longitudinal and transverse spatial directions, respectively. The metric components are independent of the boundary coordinates and depend only on $z$. $q_2$ represents the third spatial direction. We allow $g_{t\parallel}\neq0$, while the transverse direction is assumed not to mix with time or the longitudinal direction in this general setup. The metric may also contain additional spatial directions, including those that arise in higher-dimensional geometries, as well as spatial-spatial cross terms. Since the string embedding considered in the potential analysis varies along only one spatial direction, with the remaining coordinates held fixed, these additional terms do not enter the potential analysis. Their role in the DBI prescription will be examined at the end of Sec.\,2.2, and they will also be retained when required by the specific rotating backgrounds considered in Sec.\,3.

This metric provides a common framework for studying the longitudinal and transverse electric fields in the backgrounds of interest. In particular, it allows us to isolate the effect of the temporal-spatial mixing and examine its consequences for the two methods of determining the critical field.

%===================================================================================
\subsection{Potential analysis}

The heavy-quark potential in the holographic framework can be calculated from the classical Nambu--Goto action of a fundamental string in the dual gravitational background, following the prescription introduced in \cite{Maldacenawilson}. In studies of the holographic Schwinger effect, the open string is instead attached to a probe D3-brane located at a finite radial position in the bulk, so that the masses of the corresponding quarks remain finite. This prescription was employed in \cite{potential} in their potential analysis of the holographic Schwinger effect. We adopt the same setup to determine the critical electric field in a stationary background with temporal-spatial mixing.

We consider a fundamental string whose endpoints lie on a probe D3-brane at $z=z_0$ and whose profile extends into the bulk before returning to the brane. The endpoints are separated along a boundary spatial direction $q_i$, while all other spatial coordinates are
held fixed. We use the worldsheet coordinates $\tau=t$ and $\sigma=q_i$, and describe the string profile by $z=z(q_i)$. The Nambu--Goto action is
\begin{align}
S_{\mathrm{NG}}=-T_F\int d\tau\,d\sigma\,\sqrt{-\det h_{\alpha\beta}},
\label{generalNG}
\end{align}
where $T_F=(2\pi\alpha')^{-1}$ is the fundamental string tension and $h_{\alpha\beta}$ is the induced worldsheet metric. For the embedding under consideration, its components are
\begin{align}
h_{\tau\tau}&=g_{tt},\nonumber\\
h_{\tau\sigma}&=g_{ti},\nonumber\\
h_{\sigma\sigma}&=g_{ii}+g_{zz}\left(\frac{dz}{dq_i}\right)^2.
\label{generalinduced}
\end{align}
Consequently, by introducing
\begin{align}
A(z)&\equiv -g_{tt}(z)g_{zz}(z),\label{Ageneral}\\
B_i(z)&\equiv g_{ti}(z)^2-g_{tt}(z)g_{ii}(z).\label{Bgeneral}
\end{align}
the Nambu--Goto action per unit coordinate time is determined by the following effective Lagrangian density
\begin{align}
\mathcal{L}_{\mathrm{NG}}=T_F\sqrt{B_i(z)+A(z)\left(\frac{dz}{dq_i}\right)^2}.
\label{eq:NG_effective_lagrangian}
\end{align}
Since this expression does not depend explicitly on $q_i$, the corresponding Hamiltonian is conserved along the string. Let $z=z_c$ denote the turning point of the string, where
$dz/dq_i=0$. The conserved quantity can then be written as
\begin{align}
\frac{B_i(z)}{\sqrt{B_i(z)+A(z)\left(\dfrac{dz}{dq_i}\right)^2}}=\sqrt{B_i(z_c)}.
\label{generalconservedH}
\end{align}
The separation between the string endpoints is therefore
\begin{align}
L_i=2\int_{z_0}^{z_c} dz\,\sqrt{\frac{A(z)B_i(z_c)}{B_i(z)\left[B_i(z)-B_i(z_c)\right]}}.
\label{generalseparation}
\end{align}
Similarly, the energy of the connected string configuration, which gives the quark-antiquark potential in this setup, is
\begin{align}
V_{Q\bar Q}^{i}(L_i)=2T_F\int_{z_0}^{z_c} dz\,\sqrt{\frac{A(z)B_i(z)}{B_i(z)-B_i(z_c)}}.
\label{generalQpotential}
\end{align}
Since the probe brane is at a finite radial position, the string energy is finite. 

In the presence of an external electric field directed along $q_i$, the total potential becomes
\begin{align}
V_{\mathrm{tot}}^{i}(L_i;E_i)=V_{Q\bar Q}^{i}(L_i)-E_iL_i,
\label{generaltotalpotential}
\end{align}
where $E_i$ is the coordinate electric-field component conjugate to the coordinate separation $L_i$. The critical field is the threshold at which the potential barrier for quark-antiquark pair production disappears. For fields below this threshold, the total potential exhibits a barrier and pair production can occur through quantum tunneling, whereas above it there is no potential barrier preventing the production of the corresponding quark pairs.

The critical field can be determined by considering the limit $L_i\rightarrow0$, in which the string becomes increasingly short and its turning point approaches the probe brane,
$z_c\rightarrow z_0$. In this limit, the string energy satisfies
\begin{align}
E_{c,\mathrm{pot}}^i&=\lim_{L_i\rightarrow0}\frac{V_{Q\bar Q}^{i}(L_i)}{L_i}
\nonumber\\
&=T_F\sqrt{B_i(z_0)}=T_F\sqrt{g_{ti}(z_0)^2-g_{tt}(z_0)g_{ii}(z_0)}.
\label{Ecpotgeneral}
\end{align}
Thus, the critical field is determined by the metric components evaluated at the position of the probe brane.
For the longitudinal direction, $q_i=q_\parallel$, the metric component $g_{t\parallel}$ is nonzero. Equation \eqref{Ecpotgeneral} therefore gives
\begin{align}
E_{c,\mathrm{pot}}^\parallel =T_F\sqrt{ g_{t\parallel}(z_0)^2
-g_{tt}(z_0)g_{\parallel\parallel}(z_0)}.
\label{EcLpot}
\end{align}
The temporal-spatial cross term consequently contributes to the critical field expressed in terms of the longitudinal coordinate electric field.
For the transverse direction, $q_i=q_\perp$, we assume $g_{t\perp}=0$. The corresponding critical field is
\begin{align}
E_{c,\mathrm{pot}}^\perp=T_F\sqrt{-g_{tt}(z_0)g_{\perp\perp}(z_0)}.
\label{EcTpot}
\end{align}
In this direction, the critical field takes the familiar form obtained when the temporal and spatial directions do not mix.

Equations \eqref{EcLpot} and \eqref{EcTpot} provide the critical fields against which we compare the results obtained from the DBI reality condition in Sec.\,2.2.

%===================================================================================
\subsection{DBI reality condition}

An independent way to determine the critical electric field is provided by the Dirac--Born--Infeld (DBI) action of a probe D3-brane. In a diagonal background, the usual procedure is to impose the reality condition on the DBI determinant in the presence of an external electric field. For an electric field directed along a spatial coordinate $q_i$, this condition gives $E_{c,i}^{\mathrm{diag}} =T_F\sqrt{-g_{tt}(z_0)g_{ii}(z_0)}$ where the metric components are evaluated at the position $z=z_0$ of the probe brane. This result is in good agreement with that found from the potential analysis. For a stationary metric like Eq.\,\eqref{generalmetric}, however, the nonzero temporal-spatial component $g_{t\parallel}$ makes the calculation more subtle. We now examine whether the reality condition of the DBI action, in such geometries, reproduces the critical electric fields obtained from the potential analysis.   

We start from the DBI action of a probe D3-brane,
\begin{align}
S_{\mathrm{DBI}} =-T_{\mathrm{D3}}\int d^4\xi\, e^{-\phi}
\sqrt{-\det\left(P[g]_{ab}+2\pi\alpha' F_{ab}\right)},
\label{generalDBIaction}
\end{align}
where $T_{\mathrm{D3}}$ is the D3-brane tension, $P[g]_{ab}$ is the induced metric on the brane, $\phi$ is the dilaton, and $F_{ab}$ is the worldvolume gauge-field strength. The probe D3 brane fills the four space-time directions of the field theory and sits at a fixed radial position $z=z_0$. An external electric field $E$ is introduced along one of the spatial directions through the worldvolume gauge-field strength on the brane. For convenience, we define
\begin{align}
\mathcal{E}_i\equiv 2\pi\alpha' E_i =\frac{E_i}{T_F},
\label{rescaledE}
\end{align}
where $E_i=F_{ti}$ is the corresponding coordinate electric-field component. %The determinant in the DBI action must be evaluated over all worldvolume directions, including those along which the electric field vanishes. 
We first consider the metric \eqref{generalmetric} without spatial-spatial cross terms. The induced metric on the probe brane nevertheless remains non-diagonal because $g_{t\parallel}\neq0$. In particular, this component is retained irrespective of the direction of the applied electric field.

We begin with an electric field directed along $q_\parallel$, so that $F_{t\parallel}=E_\parallel$. The DBI action depends on the matrix
\begin{align}
\mathcal{M}\equiv P[g]+2\pi\alpha'F.
\end{align}
Using the worldvolume coordinates $(t,q_\parallel,q_\perp,q_2)$, where $q_2$ denotes the remaining transverse spatial direction of the D3-brane, the full worldvolume matrix entering the DBI determinant is
\begin{align}
\mathcal{M}=
\begin{pmatrix}
g_{tt} & g_{t\parallel}+\mathcal{E}_\parallel & 0 & 0\\
g_{t\parallel}-\mathcal{E}_\parallel &g_{\parallel\parallel}&0&0\\
0&0&g_{\perp\perp}&0\\ 0&0&0&g_{22}
\end{pmatrix}.
\label{DBIcoordmatrixL}
\end{align}
Here, all metric components are evaluated at $z=z_0$. The two transverse directions decouple from the $(t,q_\parallel)$ block, and the determinant becomes
\begin{align}
\det\left(\mathcal{M}\right) ={}&g_{\perp\perp}g_{22} \Big[ g_{tt} g_{\parallel\parallel} -g_{t\parallel}^2 +\mathcal{E}_\parallel^2 \Big].
\label{DBIcoorddetL}
\end{align}
The DBI reality condition requires 
\begin{align}
E_{c,\mathrm{DBI}}^{\parallel,\mathrm{coord}} =T_F\sqrt{ g_{t\parallel}(z_0)^2
-g_{tt}(z_0)g_{\parallel\parallel}(z_0) }.
\label{EcLcoord}
\end{align}
This result agrees exactly with the critical longitudinal field obtained from the potential analysis, Eq.\,\eqref{EcLpot}. Thus, in the longitudinal case, the usual DBI reality-condition calculation gives the correct result even though the background metric
contains a temporal-spatial cross term.

We next consider an electric field directed along $q_\perp$, with $F_{t\perp}=E_\perp$. Although the electric field now points along the transverse coordinate, the metric component
$g_{t\parallel}$ remains and in the same worldvolume coordinates, the relevant matrix of the brane worldvolume is
\begin{align}
\mathcal{M} = 
\begin{pmatrix}
g_{tt} & g_{t\parallel} & \mathcal{E}_\perp & 0\\
g_{t\parallel} & g_{\parallel\parallel} & 0 & 0\\
-\mathcal{E}_\perp & 0 & g_{\perp\perp} & 0\\ 0&0&0&g_{22}
\end{pmatrix}.
\label{DBIcoordmatrixT}
\end{align}
In this case, only the $q_2$ direction decouples. The remaining $(t,q_\parallel,q_\perp)$ block is three-dimensional, and its determinant is
\begin{align}
\det\mathcal{M}_{(t,\parallel,\perp)} ={}g_{\perp\perp} \left( g_{tt}g_{\parallel\parallel}
-g_{t\parallel}^2 \right)+g_{\parallel\parallel}\mathcal{E}_\perp^2.
\label{DBIcoorddetT}
\end{align}
Setting this expression to zero gives the critical transverse electric field
\begin{align}
E_{c,\mathrm{DBI}}^{\perp,\mathrm{coord}} =T_F\sqrt{ \frac{g_{\perp\perp}(z_0)} {g_{\parallel\parallel}(z_0)} \left[ g_{t\parallel}(z_0)^2 -g_{tt}(z_0)g_{\parallel\parallel}(z_0) \right] }.
\label{EcTcoord}
\end{align}
As can be seen, this expression depends explicitly on $g_{t\parallel}$, even though the applied electric field points along $q_\perp$. This result is, however, generally different from the critical transverse field obtained from the potential analysis, as given in Eq.\,\eqref{EcTpot}. This discrepancy originates from a nonzero temporal-spatial cross term in the background metric. Hence, we have demonstrated that, for a stationary geometry with a nonzero temporal-spatial term, the direct DBI reality-condition calculation reproduces the potential analysis result for a longitudinal electric field but leads to a different critical field for a transverse electric field.

The problem can become more pronounced when the metric also contains spatial-spatial cross terms. To illustrate this point, suppose that $g_{\parallel\perp}\neq0$, while retaining
$g_{t\parallel}\neq0$. Even for an electric field along $q_\parallel$, the $q_\perp$ direction no longer decouples from the $(t,q_\parallel)$ block. After the remaining transverse
direction $q_2$ has been factored out, the relevant matrix is
\begin{align}
\mathcal{M}_{(t,\parallel,\perp)}=
\begin{pmatrix}
g_{tt} & g_{t\parallel}+\mathcal{E}_\parallel & 0\\ g_{t\parallel}-\mathcal{E}_\parallel
&g_{\parallel\parallel}&g_{\parallel\perp}\\ 0&g_{\parallel\perp}&g_{\perp\perp}
\end{pmatrix}.
\label{DBIcoordL_spatialmixing}
\end{align}
Its determinant is
\begin{align}
\det\mathcal{M}_{(t,\parallel,\perp)} ={}g_{tt} \left( g_{\parallel\parallel}g_{\perp\perp}
-g_{\parallel\perp}^2 \right) -g_{\perp\perp} \left( g_{t\parallel}^2-\mathcal{E}_\parallel^2 \right).
\label{DBIcoorddetL_spatialmixing}
\end{align}
The resulting critical field is
\begin{align}
E_{c,\mathrm{DBI}}^{\parallel,\mathrm{coord}} =T_F\sqrt{ g_{t\parallel}(z_0)^2 -g_{tt}(z_0)g_{\parallel\parallel}(z_0)
+\frac{g_{tt}(z_0)g_{\parallel\perp}(z_0)^2} {g_{\perp\perp}(z_0)} }.
\label{EcLcoordspatialmixing}
\end{align}
This no longer agrees, in general, with the longitudinal critical field obtained from the potential analysis, Eq.\,\eqref{EcLpot}. The spatial cross term introduces an additional contribution to the DBI result, even though the string configuration used in the potential
analysis is oriented along $q_\parallel$ with $q_\perp$ held fixed.

These results show that the direct application of the DBI reality condition to the coordinate components of the electric field does not, in general, reproduce the potential-analysis
result in stationary non-diagonal geometries. With no spatial-spatial cross terms, the longitudinal field is recovered correctly, whereas the transverse field is not. Once spatial
cross terms are allowed, even the longitudinal result differs. We must therefore identify the origin of this discrepancy and formulate a prescription that treats the physical electric-field direction consistently with the non-diagonal geometry.

This discrepancy arises from how the electric field is specified in a stationary non-diagonal geometry. The origin of the problem is that the metric is written in a non-orthogonal
coordinate basis. The term $g_{t\parallel}$, for instance, encodes the non-orthogonality of the coordinate time and spatial directions. Consequently, a coordinate component such as $F_{t i}$ does not, in general, represent the physical electric field measured along the direction $q_i$ by a given observer. A field that is purely electric in the observer's local frame can therefore acquire additional components when expressed in the coordinate basis.

We resolve this issue by defining the external electric field in the local orthonormal frame of the relevant observer and then transforming its full field-strength tensor to the coordinate basis before substituting it into the DBI action. We will show that the resulting DBI reality condition reproduces the critical fields obtained independently from the Wilson-loop potential analysis.

We consider the stationary metric \eqref{generalmetric} and work in a region where $g_{tt}<0$. We assume no spatial-spatial cross terms, for simplicity. We choose the static observer, whose spatial coordinates remain fixed. The observer's four-velocity is
\begin{align}
u_{\rm st}^{\mu} = \frac{1}{\sqrt{-g_{tt}}} (\partial_t)^\mu .
\end{align}
An orthonormal coframe adapted to this observer is
\begin{align}
e^{\hat 0} &= \sqrt{-g_{tt}} \left( dt+\frac{g_{t\parallel}}{g_{tt}}dq_\parallel \right), \nonumber\\
e^{\hat\parallel} &= \sqrt{ g_{\parallel\parallel} -\frac{g_{t\parallel}^{\,2}}{g_{tt}}
}\,dq_\parallel,\nonumber \\
e^{\hat\perp} &=\sqrt{g_{\perp\perp}}\,dq_\perp,\nonumber\\ e^{\hat 2}&=\sqrt{g_{22}}\,dq_2.
\label{coframegeneral}
\end{align}
Here $q_2$ denotes the remaining spatial direction on the D3-brane, and all metric components are evaluated at the position of the probe brane. In this frame, the metric induced on the brane takes the Minkowski form
\begin{align}
ds_{\rm brane}^{2}=-(e^{\hat 0})^2+(e^{\hat\parallel})^2+(e^{\hat\perp})^2
+(e^{\hat 2})^2 .
\end{align}
We now apply this prescription to electric fields oriented along the transverse and longitudinal directions in the local frame.

We first consider an external electric field of physical magnitude $ E_{\rm phys}$ directed along the orthonormal direction $\hat\perp$. Its field-strength two-form is
\begin{align}
F= E_{\rm phys}\, e^{\hat 0}\wedge e^{\hat\perp}.
\end{align}
Using the coframe in Eq.\,\eqref{coframegeneral}, we obtain the coordinate-basis components
\begin{align}
F_{t\perp} = E_{\rm phys}\sqrt{-g_{tt}g_{\perp\perp}}, \qquad
F_{\parallel\perp} =- E_{\rm phys}\frac{g_{t\parallel}\sqrt{g_{\perp\perp}}}
{\sqrt{-g_{tt}}}.
\label{Finorth}
\end{align}
Thus, although the field is purely electric in the observer's orthonormal frame, its coordinate-basis representation contains both $F_{t\perp}$ and $F_{\parallel\perp}$. The second component is not an independently applied magnetic field; it is required for the coordinate-basis tensor to represent the same physical field.

Introducing 
\begin{align}
\mathcal E_{\rm phys} \equiv \frac{E_{\rm phys}}{T_F}, 
\label{MathcalE}
\end{align}
the full worldvolume matrix, in the coordinate order $(t,q_\parallel,q_\perp,q_2)$, becomes
\begin{align}
\mathcal M={}&
\begin{pmatrix}
g_{tt} & g_{t\parallel} & \mathcal E_{\rm phys}\sqrt{-g_{tt}g_{\perp\perp}} & 0 \\[4pt]
g_{t\parallel} & g_{\parallel\parallel} & -\mathcal E_{\rm phys}\displaystyle \frac{g_{t\parallel}\sqrt{g_{\perp\perp}}} {\sqrt{-g_{tt}}} & 0 \\[8pt]
-\mathcal E_{\rm phys}\sqrt{-g_{tt}g_{\perp\perp}}
& \mathcal E_{\rm phys}\displaystyle \frac{g_{t\parallel}\sqrt{g_{\perp\perp}}}
{\sqrt{-g_{tt}}} & g_{\perp\perp} & 0 \\[8pt] 0&0&0&g_{22}
\end{pmatrix}.
\label{eq:DBI_matrix_transverse_orthonormal}
\end{align}
Its determinant is
\begin{align}
\det \mathcal M= g_{22}g_{\perp\perp} \left( g_{tt}g_{\parallel\parallel} -g_{t\parallel}^{\,2} \right)
(1-\mathcal E_{\rm phys}^2).
\label{eq:DBI_det_transverse_orthonormal}
\end{align}
The metric-dependent prefactor is independent of the electric field. Using the DBI reality condition, the critical value is attained at $\mathcal E_{\rm phys,c}^2= 1$. Hence the critical
electric field measured in the chosen local orthonormal frame is
\begin{align}
E_{\rm phys,c}^{\perp}=T_F.
\label{eq:Ec_phys_transverse}
\end{align}
To compare this result with the potential analysis, we express the critical field in the coordinate convention used there. From the relation between $F_{t\perp}$ and $E_{\rm phys}$ in Eq.\,\eqref{Finorth}, we find
\begin{align}
E_{c,\mathrm{DBI}}^{\perp,\mathrm{orth}} = T_F\sqrt{-g_{tt}(z_0)g_{\perp\perp}(z_0)},
\label{EcTorth}
\end{align}
which agrees exactly with \eqref{EcTpot}. The discrepancy found using the usual coordinate-field prescription is thus removed once the electric field is defined in the observer's orthonormal frame and its full field-strength tensor is used in the DBI determinant.

We next consider a field of physical magnitude $E_{\rm phys}$ directed along $\hat\parallel$. In this case, the field-strength two-form is
\begin{align}
F =E_{\rm phys}\, e^{\hat 0}\wedge e^{\hat\parallel}.
\end{align}
Because both coframe one-forms contain only $dt$ and $dq_\parallel$, %their wedge product has no contribution proportional to $dq_\parallel\wedge dq_\parallel$. Consequently,
the only nonzero coordinate-basis component is
\begin{align}
F_{t\parallel}=E_{\rm phys} \sqrt{g_{t\parallel}^{\,2}-g_{tt}g_{\parallel\parallel}}.
\label{FtLorthonormal}
\end{align}
In contrast to the transverse case, no additional spatial component of the field strength is generated.
Using the same definition of $\mathcal E_{\rm phys}$ as in Eq.\,\eqref{MathcalE}, the DBI matrix is
\begin{align}
\mathcal M={}&
\begin{pmatrix}
g_{tt}&g_{t\parallel}&\mathcal E_{\rm phys}\sqrt{g_{t\parallel}^{\,2}-g_{tt}g_{\parallel\parallel}}&0
\\[4pt]
g_{t\parallel}&g_{\parallel\parallel}&0&0
\\[4pt]
-\mathcal E_{\rm phys}\sqrt{g_{t\parallel}^{\,2}-g_{tt}g_{\parallel\parallel}}&0&g_{\perp\perp}&0
\\[4pt]
0&0&0&g_{22}
\end{pmatrix}.
\label{DBImatrixLorthonormal}
\end{align}
Its determinant is
\begin{align}
\det \mathcal M=g_{\perp\perp}g_{22}\left(g_{tt}g_{\parallel\parallel}-g_{t\parallel}^{\,2}
\right)(1-\mathcal E_{\rm phys}^2).
\label{DBIdetLorthonormal}
\end{align}
The reality condition again gives $E_{\rm phys,c}^{\parallel}=T_F$. Using
Eq.\,\eqref{FtLorthonormal}, the corresponding coordinate critical field is
\begin{align}
E_{c,\mathrm{DBI}}^{\parallel,\mathrm{orth}}=T_F\sqrt{g_{t\parallel}(z_0)^{\,2}-g_{tt}(z_0)g_{\parallel\parallel}(z_0)},
\label{EcLorth}
\end{align}
in agreement with the potential-analysis result \eqref{EcLpot}.

We now examine whether the same prescription remains valid when the spatial part of the metric also contains off-diagonal terms. For this purpose, we extend the metric considered in Eq.\,\eqref{generalmetric} by allowing a nonzero spatial cross term $2 g_{\parallel\perp}(z)dq_\parallel dq_\perp$, while retaining the assumption that all metric components can depend only on the holographic coordinate $z$. As shown in the discussion above, introducing the electric field through only the coordinate component $F_{t\parallel}$ does not, in general, give the correct critical field. In particular, the coordinate-field prescription leads to Eq.\,\eqref{EcLcoordspatialmixing}, which contains an additional contribution proportional to $g_{\parallel\perp}^2$ and therefore does not reproduce the longitudinal result obtained from the potential analysis. We now show that this discrepancy is again removed by defining the electric field in the local orthonormal frame.

To construct the appropriate coframe, we first complete the square in the temporal part of the brane metric and obtain the orthonormal one-forms
\begin{align}
e^{\hat 0}&=\sqrt{-g_{tt}}\left(dt+\frac{g_{t\parallel}}{g_{tt}}dq_\parallel\right),
\nonumber\\e^{\hat\parallel}&=\sqrt{h_{\parallel\parallel}}\left(dq_\parallel+
\frac{h_{\parallel\perp}}{h_{\parallel\parallel}}dq_\perp\right),\nonumber
\\e^{\hat\perp}&=\sqrt{h_{\perp\perp}-\frac{h_{\parallel\perp}^2}{h_{\parallel\parallel}}}
\,dq_\perp,\nonumber\\e^{\hat 2}&=\sqrt{g_{22}}\,dq_2,
\end{align}
where
\begin{align}
h_{\parallel\parallel}=g_{\parallel\parallel}-\frac{g_{t\parallel}^2}{g_{tt}},
\qquad h_{\perp\perp}=g_{\perp\perp}, \qquad h_{\parallel\perp}=g_{\parallel\perp}.
\end{align}
Here we work in a region where $g_{tt}<0$ and the spatial metric defined by $h_{ij}$ is positive definite.

We consider an electric field of physical magnitude $E_{\rm phys}$ directed along the orthonormal direction $\hat\parallel$. Its field-strength two-form is
\begin{align}
F=E_{\rm phys}\,e^{\hat 0}\wedge e^{\hat\parallel}.
\end{align}
Unlike the case with no spatial-spatial cross term, the coordinate-basis representation of this field contains additional components. Explicitly, the coordinate-basis components of the field-strength are
\begin{align}
F_{t\parallel}=E_{\rm phys}\sqrt{-g_{tt}h_{\parallel\parallel}}
, \qquad F_{t\perp}=E_{\rm phys}\sqrt{-g_{tt}h_{\parallel\parallel}}\,\frac{h_{\parallel\perp}}{h_{\parallel\parallel}}
,\qquad F_{\parallel\perp}=E_{\rm phys}\sqrt{-g_{tt}h_{\parallel\parallel}}\,\frac{g_{t\parallel}}{g_{tt}}\frac{h_{\parallel\perp}}{h_{\parallel\parallel}}.
\end{align}
Then, the determinant of the DBI matrix becomes
\begin{align}
\det M=g_{22}g_{tt}\left(h_{\parallel\parallel}h_{\perp\perp}
-h_{\parallel\perp}^2\right)\left(1-\mathcal E_{\rm phys}^2\right).
\end{align}
The DBI reality condition therefore again gives $E_{\rm phys,c}^{\parallel}=T_F$. The coordinate component of the field strength conjugate to the separation along $q_\parallel$ is
\begin{align}
F_{t\parallel}=E_{\rm phys}\sqrt{-g_{tt}h_{\parallel\parallel}}
=E_{\rm phys}\sqrt{g_{t\parallel}^2-g_{tt}g_{\parallel\parallel}}.
\end{align}
Consequently, the corresponding coordinate critical field is
\begin{align}
E_{c,\mathrm{DBI}}^{\parallel,\mathrm{orth}}
=T_F\sqrt{g_{t\parallel}(z_0)^2-g_{tt}(z_0)g_{\parallel\parallel}(z_0)},
\end{align}
which is independent of $g_{\parallel\perp}$ and agrees with the longitudinal critical field obtained from the potential analysis, Eq.\,\eqref{EcLpot}.
This result shows that the orthonormal-frame prescription continues to reproduce the potential-analysis result even in the presence of spatial-spatial mixing. In contrast, the
coordinate-field prescription does not, in general, yield the correct longitudinal critical field in this case.

We have therefore shown that the DBI reality condition reproduces the critical fields obtained from the Wilson-loop potential analysis when the applied electric field is defined in the local
orthonormal frame of the observer and its full field-strength tensor is transformed to the coordinate basis. This prescription works not only for the stationary metric considered initially, but also in the presence of spatial-spatial cross terms, for which the coordinate-field prescription can fail even in the longitudinal direction. In the next section, we apply the same prescription to explicit rotating geometries, including the global Myers--Perry background, to further illustrate its implementation in geometries with nontrivial temporal-spatial and spatial-spatial mixing.

%*************************************************************************
\section{Critical electric field in rotating Myers--Perry backgrounds}

In this section, we investigate the critical electric field in two rotating Myers--Perry--AdS$_5$ backgrounds \cite{MPBH,HawkingMP}: the global equal-rotation black hole and its planar limit. We first introduce the global geometry and then obtain the planar-limit metric by taking the large-horizon, small-angle limit. This establishes the relation between the two backgrounds before we perform the potential analysis and the DBI calculation in each case.

The Myers--Perry (MP) black hole is a full spinning asymptotically global AdS solution of the Einstein equations with two independent conserved angular momenta $a$ and $b$. The line element of the five-dimensional MP black hole with equal angular momenta $b=a$ can be written as
\begin{align}
ds^2={}&-dt^2+\frac{R^4}{z^4G(z)}\,dz^2+\frac{R^2}{z^2}\,ds_{\rm bdy}^2
+\frac{2\mu z^2}{R^4}\left[dt+\frac{a}{2}\left(d\psi+\cos\theta\,d\phi\right)\right]^2,
\label{metricMPglobal}
\end{align}
where 
\begin{align}
ds_{\rm bdy}^2={}&-dt^2+\frac{R^2}{4}\left[d\theta^2+d\psi^2+d\phi^2
+2\cos\theta\,d\psi\,d\phi\right],
\label{metricbdyMPglobal}
\end{align}
is the line element of the conformal boundary which has a compact topology $\mathbb{R}\times S^3$, with the spatial part describing a three-sphere of radius $R/2$ in Hopf coordinates. The angular coordinates are in the ranges $0\leq\theta\leq\pi$, $0\leq\psi<4\pi$, and $0\leq\phi<2\pi$. With this choice of coordinates, the boundary is static and the rotation occurs along the $\psi$ direction \cite{MPinpsi1,MPinpsi2,Braga2026}. Hence, the longitudinal direction is chosen to be $\psi$, which is the direction associated with the rotation, while $\theta$ is taken as the transverse direction (see \cite{frameus}). The blackening function is
\begin{align}
G(z)={}&1+\frac{R^2}{z^2}-\frac{2\mu z^2}{R^4}
\left(1-\frac{a^2}{R^2}\right)+\frac{2\mu a^2z^4}{R^8}.
\label{MPG}
\end{align}
The conformal boundary is located at $z=0$, while the horizon is
located at $z=z_h$, where $G(z_h)=0$. Solving this condition gives
the mass parameter in terms of the horizon position:
\begin{align}
\mu=\frac{R^8(R^2+z_h^2)}{2z_h^4\left[R^2(R^2-a^2)-a^2z_h^2\right]}.
\label{MPmu}
\end{align}
Moreover, the Hawking temperature of the black hole is given by
\begin{align}
T_{\rm global}=\frac{2R^6+R^4z_h^2-2a^2(R^2+z_h^2)^2}{2\pi R^4z_h
\sqrt{R^4-a^2(R^2+z_h^2)}}.
\label{globaltemp}
\end{align}

The planar-limit geometry is obtained by expanding the global solution around a point on the boundary three-sphere and taking the large-horizon limit. To identify the direction associated
with the rotation, we choose the coordinate replacements
\begin{align}
\frac{R}{2}(\psi-2\pi)&=\varepsilon x, \\
\frac{R}{2}(\phi-\pi)&=\varepsilon y_1, \\
\frac{R}{2}\tan\left(\theta-\frac{\pi}{2}\right) &=\varepsilon y_2,
\end{align}
together with the scalings
\begin{align}
t\longrightarrow\varepsilon t,\qquad z\longrightarrow\varepsilon z,\qquad
z_h\longrightarrow\varepsilon z_h.
\label{planarscaling}
\end{align}
Applying these replacements to the metric of the global MP rotating given in Eq.\,\eqref{metricMPglobal} and taking the limit $\varepsilon\longrightarrow0$, the metric becomes
\begin{align}
ds^2=\frac{R^2}{z^2}\left[-dt^2+dx^2+dy_1^2+dy_2^2+\frac{z^4}{z_h^4\left(1-a^2/R^2\right)} \left(dt+\frac{a}{R}\,dx\right)^2+\frac{dz^2}{1-z^4/z_h^4}\right].
\label{MPplanarmetric}
\end{align}
The coordinate $z$ is again the bulk radial coordinate. The boundary is at $z=0$, and the horizon is at $z=z_h$. The only time-space cross term is $g_{tx}$, so the $x$ direction is distinguished from the transverse directions $y_1$ and $y_2$, as the longitudinal Cartesian coordinate along the rotation-induced flow. For $a=0$, the metric reduces to the planar Schwarzschild--AdS$_5$ black-brane geometry.

The Hawking temperature of the planar-limit geometry is
\begin{align}
T_{\rm planar}=\frac{1}{\pi z_h} \sqrt{1-\frac{a^2}{R^2}}.
\label{planartemp}
\end{align}
Unlike the global solution, this limiting geometry has a flat boundary and describes a boosted black brane.

The thermodynamic phase structure of the global background is determined by the Hawking--Page transition, which corresponds to the equality of the free energies of the rotating black hole and thermal AdS \cite{HPT}. Following Witten's proposal \cite{Witten1998HP}, this transition is interpreted as the confinement/deconfinement transition of the dual gauge theory. The corresponding critical temperature $T_{\rm HP}$ depends on the rotation parameter and is obtained from the thermodynamic analysis of the solution, for example by comparing the on-shell actions of the rotating black-hole and thermal-AdS solutions. For the global Myers–Perry background considered here, this analysis has been carried out in \cite{Braga2026,frameus}; see also \cite{HawkingMP,thermokerr} for earlier studies of rotating AdS black-hole thermodynamics. In the numerical analysis below, the temperature of the global black-hole background is chosen above $T_{\rm HP}$, so that the black hole is thermodynamically favored and the dual system is in the deconfined phase.

In the planar limit, the boundary spatial volume becomes noncompact, and the finite-temperature Hawking--Page transition is absent. The planar black-brane geometry therefore describes the deconfined phase for any nonzero Hawking temperature, and no finite critical temperature needs to be imposed in this case.

In the next subsection, we calculate the critical electric field in the planar-limit geometry using the Wilson-loop potential analysis and the DBI reality condition. Then, we perform the corresponding analysis in the global geometry, retaining its angular structure and the relevant spatial cross terms. Comparing these results will allow us to test the prescription
developed in Sec.\,2 in both geometries.

%=======================================================
\subsection{Planar-limit rotating black-hole geometry}

We first consider the planar-limit geometry associated with the rotating Myers--Perry AdS$_5$ black hole. We begin by specifying the admissible parameter range and the
worldsheet structure relevant to the quark--antiquark configurations. Throughout our discussion, we take $a\geq0$, since its sign can be reversed by changing the orientation of the rotation. The planar geometry describes a boosted black brane, with boundary velocity
$v=a/R$. Causality therefore requires $a<R$. 

The position $z_0$ of the probe D3-brane, which is related to the mass of the external quarks in the dual field theory, must lie in a region where the effective string tension is real and nonzero. For strings oriented along the longitudinal and transverse directions, respectively, the effective tensions are
\begin{align}
\sigma_\parallel^{\rm planar}(z)&=T_f \sqrt{B_\parallel^{\rm planar}(z)}=T_f\sqrt{g_{tx}^2-g_{tt}g_{xx}}=T_f \frac{R^2}{z^2}\sqrt{1-\frac{z^4}{z_h^4}},
\label{eq:planar_sigma_parallel}\\
\sigma_\perp^{\rm planar}(z)&=T_f \sqrt{B_\perp^{\rm planar}}=T_f\sqrt{-g_{tt}g_{y_1y_1}}=T_f \frac{R^2}{z^2}
\sqrt{1-\frac{z^4}{z_h^4(1-a^2/R^2)}}.
\label{eq:planar_sigma_perp}
\end{align}
The longitudinal tension vanishes at the black-brane horizon $z_h$, whereas the transverse tension vanishes at
\begin{align}
z_*=z_h\left(1-\frac{a^2}{R^2}\right)^{1/4}=\frac{1}{\pi T}
\left(1-\frac{a^2}{R^2}\right)^{3/4}<z_h, \qquad a\neq0.
\label{eq:planar_zstar}
\end{align}
Notice that in the nonrotating limit we have $z_*(a=0)= z_h$. This latter surface coincides with the zero of $g_{tt}$ and acts as a worldsheet horizon for the static transverse string. The appearance of such a worldsheet horizon is a consequence of the rotating planar geometry and is familiar in holographic descriptions of moving probes and trailing strings \cite{zstar1,zstar2,zstar3,zstar4,zstar5,zstar6}, as well as other studies on rotating strongly coupled plasma in holography \cite{thermoscaling,frameus}. Requiring longitudinal (transverse) effective tension to be real and nonzero at the probe brane leads to the condition $z_0<z_h$ ($z_0<z_*$). At fixed $T$ and $z_0$, this gives $0\leq a<a_{\max}$ with
\begin{align}
a_{\max}^{\parallel}=R\sqrt{1-(\pi Tz_0)^{2}}, \qquad a_{\max}^{\perp}=R\sqrt{1-(\pi Tz_0)^{4/3}}, \qquad \pi Tz_0<1.
\label{planaramax}
\end{align}

We use $R=1\,\mathrm{GeV}^{-1}$ and $z_0=0.1\,\mathrm{GeV}^{-1}$ throughout the analysis. The probe-brane position $z_0$ is related to the mass of the external quarks, and fixing it is sufficient for the comparison of the critical-field prescriptions considered here. For the following plots, we choose the values of $a$ and $T$ so as to lie within the admissible parameter ranges.

We first determine the critical electric fields using the potential analysis and compare them with the results obtained from the DBI reality condition. Substituting the metric components from the planar geometry into the results found from the potential analysis in Eqs.\,\eqref{EcLpot} and \eqref{EcTpot}, we obtain
\begin{align}
E_{c,\mathrm{pot}}^\parallel&=\frac{T_FR^2}{z_0^2}
\sqrt{1-\frac{z_0^4}{z_h^4}},
\label{eq:planar_Ec_parallel_zh}\\
E_{c,\mathrm{pot}}^\perp&=\frac{T_FR^2}{z_0^2}
\sqrt{1-\frac{z_0^4}{z_h^4(1-a^2/R^2)}}.
\label{eq:planar_Ec_perp_zh}
\end{align}
An important feature of the longitudinal result is that the effective string tension and consequently the critical electric field is independent of the rotation parameter $a$ when the horizon position $z_h$ is held fixed. At fixed temperature, however, $z_h$ depends on
$a$ through the temperature relation given in Eq.\,\eqref{planartemp}. Consequently, it can be written as
\begin{align}
E_{c,\mathrm{pot}}^\parallel=\frac{T_FR^2}{z_0^2}
\sqrt{1-\frac{(\pi Tz_0)^4}{\left(1-a^2/R^2\right)^2}}.
\label{eq:planar_Ec_parallel_T}
\end{align}
Thus, at fixed $T$ and $z_0$, the longitudinal critical field decreases with increasing $a$. Its dependence on rotation is entirely due to the change in $z_h$ required to maintain the same temperature, rather than to an explicit rotation dependence of the longitudinal effective
string tension. In contrast, the transverse effective string tension depends explicitly on $a$ as well as on $z_h$.

We now compare these expressions with the DBI results. Using the orthonormal-frame prescription, when the electric field is specified in the local orthonormal frame and the full field-strength tensor is transformed to the coordinate basis, the DBI reality condition gives the same results as the potential analysis for both directions. The coordinate-field prescription, in which the electric field is introduced by assuming that only the corresponding time--space component of the coordinate-basis field-strength tensor is nonzero, gives the same longitudinal critical field as described in Sec.\,2.2. In the transverse direction, however, it gives 
\begin{align}
E_{c,\mathrm{DBI}}^{\perp,\mathrm{coord}}
=T_F\sqrt{ \frac{g_{y_1 y_1}} {g_{xx}} \left[ g_{tx}^2 -g_{tt}g_{xx} \right] }\bigg|_{z=z_0}
=\frac{T_FR^2}{z_0^2}\sqrt{\frac{\left(a^2-R^2 \right)\left(z_0^4-z_h^4\right)}{R^2 z_h^4+a^2\left(z_0^4-z_h^4\right)}}.
\label{EcTcoordplanar}
\end{align}
This obviously differs from $E_{c,\mathrm{pot}}^\perp=E_{c,\mathrm{DBI}}^{\perp,\mathrm{orth}}$, whereas the two transverse results coincide for $a=0$.

This issue is relevant to previous studies of the holographic Schwinger effect in rotating backgrounds. In particular, the transverse critical field obtained in \cite{rotSch3} employs the coordinate-field prescription and therefore differs from the result obtained here using the physical orthonormal-frame field. This comparison illustrates why the identification of the applied electric field requires particular care in a stationary, non-diagonal geometry.

\begin{figure}[t]
\centering
\includegraphics[width=0.48\textwidth]{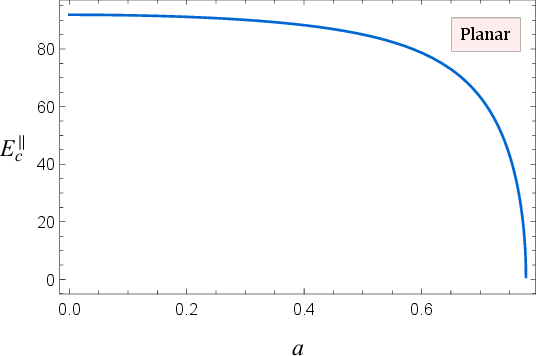}
\includegraphics[width=0.48\textwidth]{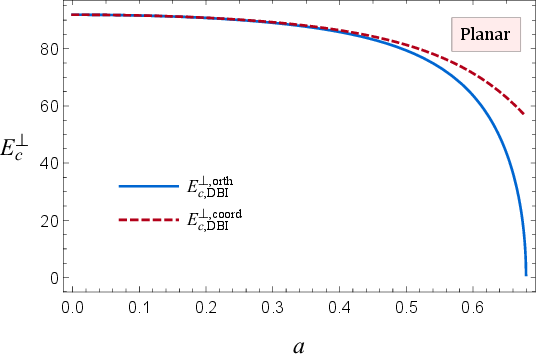}
\caption{\footnotesize Critical electric fields as functions of the rotation parameter $a$ at fixed $T=2\,\mathrm{GeV}$ and $z_0=0.1\,\mathrm{GeV}^{-1}$ in the planar-limit geometry. Left: longitudinal direction, where the potential analysis and both DBI prescriptions give the same curve. Right: transverse direction, where the potential analysis agrees with the orthonormal-frame DBI prescription, while the coordinate-field prescription yields a larger critical field for $a\neq0$. Only the physically admissible range of $a$ is shown.}
\label{Ecplanar}
\end{figure}

Figure \ref{Ecplanar} displays the critical fields as functions of $a$ at fixed $T=2\,\mathrm{GeV}$ and $z_0=0.1\,\mathrm{GeV}^{-1}$. In the longitudinal panel, all three
methods give the same result and are therefore represented by a single curve. In the transverse panel, the potential analysis and the orthonormal-frame DBI prescription likewise produce a single common curve, while the coordinate-field DBI result lies above it for
nonzero $a$. The coordinate-field prescription therefore overestimates the transverse critical field. The separation between the two transverse curves increases with $a$. In both panels, the curves are restricted to the physically admissible range determined by Eq.\,\eqref{planaramax}. The critical electric fields in both directions decrease with $a$ until $E_{c,\mathrm{pot}}^\parallel=E_{c,\mathrm{DBI}}^{\parallel,\mathrm{orth}}$ and $E_{c,\mathrm{pot}}^\perp=E_{c,\mathrm{DBI}}^{\perp,\mathrm{orth}}$ vanish at a rotation parameter for which $z_h\to z_0$ and $z_* \to z_0$, respectively.

We next investigate the separation of the quark--antiquark pair and the corresponding total potential. We consider a connected U-shaped string with turning point $z=z_c$ and endpoints on the probe brane at $z=z_0$. The separation $L_i(z_c)$, with $i=\parallel,\perp$, and the on-shell Nambu--Goto energy are obtained from the string equations of
motion, using the full induced worldsheet metric for each orientation.
In particular, using the planar-limit metric components given in Eq.\,\eqref{MPplanarmetric} for the strings oriented in $x$ and $y_1$ directions, we have
\begin{align}
A_{\rm planar}(z)&=\frac{R^4}{z^4}\frac{R^2 z^4 -\left(R^2-a^2\right)z_h^4}{\left(R^2-a^2\right)\left(z^4-z_h^4\right)},\nonumber\\
B_\parallel^{\rm planar}(z)&=R^2\sqrt{\frac{1}{z^4}-\frac{1}{z_h^4}}, \nonumber\\ B_\perp^{\rm planar}(z)&= \frac{R^2}{z^2} \sqrt{1-\frac{z^4}{z_h^4(1-a^2/R^2)}}.
\label{planarAB}
\end{align}
Substituting these relations into the relations of the separation length and the quark-antiquark potential given in Eqs.\,\eqref{generalseparation}-\eqref{generaltotalpotential}, the separation length, $L_i$, and the total potential  in the presence of an external electric field, $V_{\mathrm{tot}}(L_i;E_i)$, for the present geometry are obtained for both directions. 

\begin{figure}[t]
\centering
\includegraphics[width=0.48\textwidth]{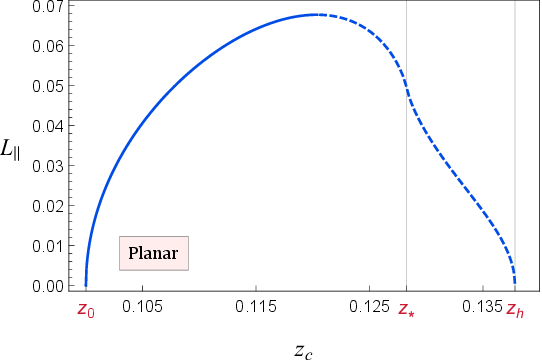}
\includegraphics[width=0.48\textwidth]{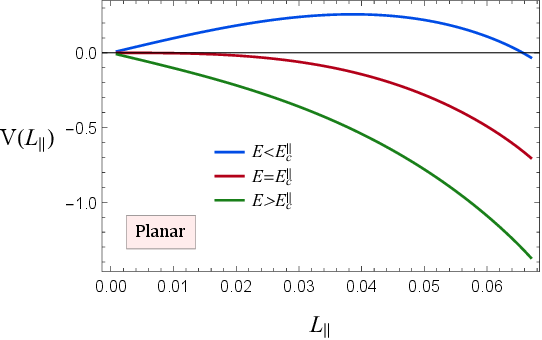}
\caption{\footnotesize Longitudinal quark--antiquark configuration in the
planar-limit geometry for $T=2\,\mathrm{GeV}$, $a=0.5\,\mathrm{GeV}^{-1}$, and $z_0=0.1\,\mathrm{GeV}^{-1}$. Left: separation $L_\parallel$ as a function of the turning point $z_c$. Right: total potential as a function of $L_\parallel$ for external fields below,
at, and above $E_c^\parallel$. The barrier disappears at the common critical field obtained from the potential analysis and both DBI prescriptions.}
\label{LVLplanar}
\end{figure}

\begin{figure}[t]
\centering
\includegraphics[width=0.48\textwidth]{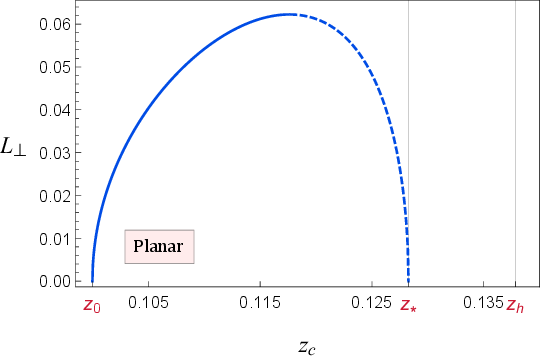}
\includegraphics[width=0.48\textwidth]{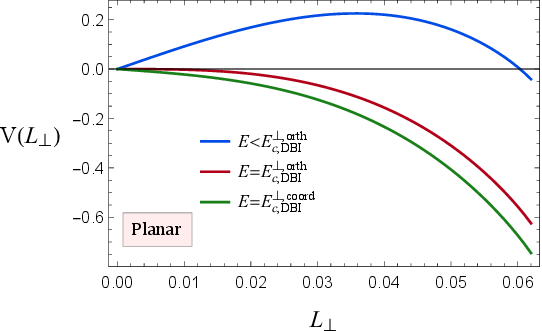}
\caption{\footnotesize Transverse quark--antiquark configuration in the
planar-limit geometry for $T=2\,\mathrm{GeV}$, $a=0.5\,\mathrm{GeV}^{-1}$, and $z_0=0.1\,\mathrm{GeV}^{-1}$. Left: separation $L_\perp$ as a function of $z_c$, showing the termination of the farther branch at the worldsheet horizon $z_*$. Right: total
potential for fields below and at the physical critical value 
$E_{c,\mathrm{DBI}}^{\perp,\mathrm{orth}}=E_{c,\mathrm{pot}}^\perp$,
as well as at the larger coordinate-field value
$E_{c,\mathrm{DBI}}^{\perp,\mathrm{coord}}$. The barrier disappears at the former value, while the potential is already monotonically decreasing at the latter.}
\label{LVTplanar}
\end{figure}

Figures \ref{LVLplanar} and \ref{LVTplanar} present the separation $L_i$ as a function of $z_c$ in their left panels and the total potential as a function of $L_i$ in their right panels. We use $T=2\,\mathrm{GeV}$, $a=0.5\,\mathrm{GeV}^{-1}$, and
$z_0=0.1\,\mathrm{GeV}^{-1}$. These parameters lie within the admissible range.

The separation curves exhibit the characteristic two-branch structure of the deconfined phase: for separations below a maximum value $L_i^{\max}$, two connected-string solutions exist. The solution with the turning point farther from the boundary is energetically disfavored relative to the other connected solution. The two orientations differ in their behavior toward the interior. For the transverse configuration, the farther branch terminates at $z=z_*<z_h$, where $g_{tt}$ vanishes. This surface acts as a worldsheet horizon for the static transverse string. For the longitudinal configuration, the effective string tension
$\sigma_\parallel^{\rm planar}(z)$ remains nonzero at $z_*$, and the separation curve exhibits a change in behavior there rather than the same termination. This difference reflects the distinct effective string tensions in the two orientations.

The right panel of Fig.\,\ref{LVLplanar} shows the longitudinal total potential for fields below, at, and above the common critical value. The barrier disappears at $E=E_c^\parallel$, consistently with the potential analysis and both DBI prescriptions.
The transverse case provides a direct test of the competing DBI prescriptions. In the right panel of Fig.\,\ref{LVTplanar}, the barrier disappears at $E=E_{c,\mathrm{DBI}}^{\perp,\mathrm{orth}}=E_{c,\mathrm{pot}}^\perp$. At the larger value $E=E_{c,\mathrm{DBI}}^{\perp,\mathrm{coord}}$, the total potential is already monotonically decreasing. Hence, the coordinate-field result does not identify the onset of barrier disappearance and cannot be interpreted as the physical transverse critical field.

%=======================================================
\subsection{Global Myers--Perry rotating black hole geometry}

We now turn to the global equal-rotation Myers--Perry AdS$_5$ black hole geometry whose boundary geometry is spherical rather than planar, introduced at the beginning of Sec.~3. In contrast to the planar limit, the global system undergoes a Hawking--Page transition between thermal AdS and the black-hole phase. Since we are interested in the deconfined plasma described by the MP black-hole solution, we restrict the parameters to the large-black-hole branch above the Hawking--Page temperature.

In terms of the rotating parameter $a$, the Hawking--Page transition occurs at \cite{Braga2026}
\begin{align}
z_{\mathrm{HP}}(a) =R\sqrt{\frac{1-a^2/R^2}{1+a^2/R^2}},
\label{global_zHP}
\end{align}
with critical temperature
\begin{align}
T_{\mathrm{HP}}(a) =\frac{3-a^2/R^2}{2\pi R(1+a^2/R^2)}.
\label{Tcglobal}
\end{align}
The critical temperature reduces to $T_{\mathrm{HP}}(0) =\frac{3}{2\pi R}$ in the absence of rotation. Accordingly, the deconfined phase considered below satisfies $T>T_{\mathrm{HP}}(a)$ with  $z_h<z_{\mathrm{HP}}(a)$, where $z_h$ is determined from the temperature relation of the global geometry in Eq.\,\eqref{globaltemp}. At fixed $T$ and $a$, this relation has two black-hole branches. We choose the small-$z_h$ root, corresponding to the thermodynamically stable large-black-hole branch; the second root corresponds to an unstable solution and does not describe the deconfined phase of interest here.

The rotation parameter must also obey the causality condition on the boundary velocity. The thermodynamic angular velocity in the boundary-static coordinates is given by \cite{Braga2026}
\begin{align}
\Omega=-2a\left(\frac{1}{R^2}+\frac{z_h^2}{R^4}\right).
\end{align}
Since the boundary three-sphere has radius $R/2$, the corresponding linear velocity is $v=|\Omega|R/2$. Requiring $v<1$ and choosing $a\geq0$ without loss of generality give
\begin{align}
0\leq a<\frac{R}{1+z_h^2/R^2}.
\label{globalvelocitybound}
\end{align}
The admissible values of $a$ at fixed temperature must therefore be determined together with the chosen large-black-hole solution $z_h(T,a)$; the velocity bound cannot in general be replaced by a constant upper limit independent of the horizon position. 

The stationary-time component $g_{tt}$ vanishes at
\begin{align}
z_*=\frac{R^2}{2\sqrt{\mu}}\sqrt{1+\sqrt{1+\frac{8\mu}{R^2}}}\,,
\label{eq:global_zstar}
\end{align}
with $z_*<z_h$. The transverse effective string tension,
$\sigma_\perp^{\rm global}(z)=T_f\sqrt{-g_{tt}g_{\theta\theta}}$, therefore vanishes at $z=z_*$, whereas the longitudinal effective string tension,
$\sigma_\parallel^{\rm global}(z)=T_f\sqrt{g_{t\psi}^2-g_{tt}g_{\psi\psi}}$, does not vanish there. Replacing the metric components shows that the longitudinal tension vanishes at the black-hole horizon $z_h$. Requiring the effective tensions to remain real and nonzero at the probe-brane position consequently imposes the stronger condition $z_0<z_*$ for the transverse configuration and $z_0<z_h$ for the longitudinal configuration. These restrictions must be imposed when choosing the parameters used in the numerical analysis.

As in the planar case, we set $R=1\,\mathrm{GeV}^{-1}$ and $z_0=0.1\,\mathrm{GeV}^{-1}$ for the displayed results, and we restrict all plots to parameter values satisfying the deconfinement, probe-brane, and boundary-velocity conditions above. We then compare
the critical electric fields obtained from the potential analysis and the two DBI prescriptions, followed by the quark--antiquark separation and total-potential curves in the longitudinal and transverse directions. This comparison will establish whether the agreement between the potential analysis and the orthonormal-frame DBI prescription, and the discrepancy associated with the coordinate-field prescription, persist in the global geometry.

%======================================================
We first determine the critical electric fields from the potential analysis. We take
$\psi$ and $\theta$ as the longitudinal and transverse directions, respectively. Using the metric components of the global rotating geometry, the metric combinations in Eq.\,\eqref{Bgeneral} can be written explicitly as
\begin{align}
B_\parallel^{\rm global}(z) &=g_{t\psi}^2-g_{tt}g_{\psi\psi}=\frac{R^6}{4z^4}+\frac{R^4}{4z^2}-\frac{\mu}{2}+\frac{\mu a^2}{2R^2}+\frac{\mu a^2z^2}{2R^4},
\label{globalBL}\\
B_\perp^{\rm global}(z)&=-g_{tt}g_{\theta\theta}=\frac{R^6}{4z^4}+\frac{R^4}{4z^2}
-\frac{\mu}{2}.
\label{globalBT}
\end{align}
Consequently, following the general expressions derived in Eq.\,\eqref{Ecpotgeneral}, the critical fields obtained from the potential analysis are
\begin{align}
E_{c,\mathrm{pot}}^\parallel&=\frac{2T_F}{R}\sqrt{B_\parallel^{\rm global}(z_0) }=\frac{2T_F}{R}\left(\frac{R^6}{4z_0^4}+\frac{R^4}{4z_0^2}-\frac{\mu}{2}+\frac{\mu a^2}{2R^2}+\frac{\mu a^2z_0^2}{2R^4}\right)^{\!1/2},
\label{globalEcLpot}\\
E_{c,\mathrm{pot}}^\perp&=\frac{2T_F}{R}\sqrt{B_\perp^{\rm global}(z_0)}=\frac{2T_F}{R}\left(\frac{R^6}{4z_0^4}+\frac{R^4}{4z_0^2}-\frac{\mu}{2}\right)^{\!1/2}.
\label{globalEcTpot}
\end{align}
The factor $2/R$ in these expressions accounts for the normalization of the angular coordinates in the boundary metric. In particular, the proper length associated with an angular displacement is measured in units of $R/2$, rather than by the coordinate
displacement itself. Thus, the critical fields in Eqs.\,\eqref{globalEcLpot} and
\eqref{globalEcTpot} are expressed in terms of the physical electric field in the boundary theory.

We now compare these results with those obtained from the DBI reality condition. As in the planar case, we first consider the prescription in which the electric field is introduced through a single nonzero coordinate component of the field strength. For a field along the longitudinal direction, this means taking $F_{t\psi}\ne0$, while for a transverse field we take
$F_{t\theta}\ne0$. In contrast to the planar geometry, the global rotating metric contains both time--space and spatial--spatial off-diagonal components. In particular, the relations
\begin{align}
g_{t\phi}=\cos\theta\,g_{t\psi}, \qquad g_{\phi\psi}=\cos\theta\,g_{\psi\psi}
\label{globalcrossrelations}
\end{align}
must be taken into account when evaluating the DBI determinant. The resulting critical fields from the coordinate-field prescription are
\begin{align}
E_{c,\mathrm{DBI}}^{\parallel,\mathrm{coord}}&=\frac{2T_F}{R}
\sqrt{\left(1-\frac{g_{\psi\psi}}{g_{\phi\phi}}\cos^2\theta\right)\left(g_{t\psi}^2-g_{tt}g_{\psi\psi}\right)}\bigg|_{z=z_0}\nonumber\\&=\frac{2T_F}{R}\sqrt{\frac{R^8\sin^2\theta}
{{R^8+2\mu a^2z_0^4\cos^2\theta}}\left(\frac{R^6}{4z_0^4}+\frac{R^4}{4z_0^2}-\frac{\mu}{2}+\frac{\mu a^2}{2R^2}+\frac{\mu a^2z_0^2}{2R^4}\right)},
\label{EcLDBIcoordglobal}\\
E_{c,\mathrm{DBI}}^{\perp,\mathrm{coord}}&=\frac{2T_F}{R}\sqrt{\frac{g_{\theta\theta}}{g_{\psi\psi}}\left(g_{t\psi}^2-g_{tt}g_{\psi\psi}\right)}\bigg|_{z=z_0}\nonumber\\&=\frac{2T_F}{R}\sqrt{\frac{R^8}{R^8+2\mu a^2z_0^4}\left(\frac{R^6}{4z_0^4}+\frac{R^4}{4z_0^2}-\frac{\mu}{2}+\frac{\mu a^2}{2R^2}+\frac{\mu a^2z_0^2}{2R^4}\right)}.
\label{EcTDBIcoordglobal}
\end{align}
The dependence on $\theta$ cancels from the transverse expression, whereas the longitudinal result retains an explicit angular dependence. At the equator, $\theta=\pi/2$, the longitudinal coordinate-field result reduces to the corresponding result from the potential analysis. Away from the equator, however, the two results do not agree, even in the nonrotating limit $a=0$. In particular, at $a=0$, the extra factor in the longitudinal expression reduces to $\sin^2\theta$. This feature reflects the fact that the coordinate component $F_{t\psi}$, with all other components set to zero, does not represent a physical electric field along the longitudinal direction in the local orthonormal frame. Even in the nonrotating geometry, the physical longitudinal direction is associated with the one-form $d\psi+\cos\theta\,d\phi$, rather than with $d\psi$ alone. Thus, the coordinate-field prescription does not describe the intended physical longitudinal field away from the equator. At $\theta=\pi/2$, the spatial cross-term $g_{\psi\phi}$ vanishes, and the coordinate prescription gives the same result as the potential analysis.

More generally, the extra factor in the longitudinal expression would become unity if all off-diagonal metric components involving the longitudinal coordinate, including $g_{\psi\phi}$, were absent. In that case, if also $g_{t\psi}=0$, the usual expression $E_c=T_F\sqrt{-g_{tt}g_{\psi\psi}}$ would be recovered, up to the appropriate normalization $R/$ associated with the boundary coordinates. The transverse coordinate-field result, on the other hand, differs from the potential result for $a\neq0$. This discrepancy arises from the temporal-spatial mixing encoded in $g_{t\psi}$ and $g_{t\phi}$: a field represented only by $F_{t\theta}$ in the coordinate basis does not correspond to a purely transverse electric field in the local orthonormal frame. 

We next apply the orthonormal-frame prescription developed in Sec.\,2. This case provides a further test of the prescription because the spatial part of the metric is not diagonal. To implement the orthonormal-frame prescription, we first construct an orthonormal coframe for the global rotating geometry. Completing the square in the time and angular coordinates, and using the relations among the off-diagonal metric components in Eq.\,\eqref{globalcrossrelations}, we obtain
\begin{align}
e^{\hat t}&=\sqrt{-g_{tt}}\left[dt+\frac{g_{t\psi}}{g_{tt}}
\left(d\psi+\cos\theta\,d\phi\right)\right],
\nonumber\\
e^{\hat\theta}&=\sqrt{g_{\theta\theta}}\,d\theta,
\nonumber\\
e^{\hat\psi}&=\sqrt{h_{\psi\psi}}\,\left(d\psi+\cos\theta\,d\phi\right),
\nonumber\\
e^{\hat\phi}&=\sqrt{g_{\theta\theta}}\sin\theta\,d\phi,
\label{globalorthcoframe}
\end{align}
where
\begin{align}
h_{\psi\psi}=g_{\psi\psi}-\frac{g_{t\psi}^2}{g_{tt}}.
\end{align}
Here, the spatial one-forms have been chosen so that the spatial part of the metric is diagonal in the orthonormal frame. In particular, the combination $d\psi+\cos\theta\,d\phi$ accounts for the spatial cross-term $g_{\psi\phi}$.

For a physical electric field $E_{\rm phys}^\parallel$ along the longitudinal direction, the field strength in this frame is
\begin{align}
F=E_{\rm phys}^\parallel\,e^{\hat t}\wedge e^{\hat\psi}.
\label{globalFL}
\end{align}
Substituting the coframe of Eq.\,\eqref{globalorthcoframe} and transforming the resulting two-form to the coordinate basis, we find the nonzero components
\begin{align}
F_{t\psi}&=E_{\rm phys}^\parallel \sqrt{g_{t\psi}^2-g_{tt}g_{\psi\psi}},
\nonumber\\
F_{t\phi}&=\cos\theta\,F_{t\psi},
\label{globalFLcomponents}
\end{align}
while $F_{\phi\psi}$ becomes zero using the relations among the off-diagonal metric components. Thus, a physical longitudinal electric field in the orthonormal frame is represented by both $F_{t\psi}$ and $F_{t\phi}$ in the coordinate basis. The additional component is required by the spatial cross-term of the metric and must be retained when
evaluating the DBI action.

For a physical electric field $E_{\rm phys}^\perp$ along the transverse direction, we instead take
\begin{align}
F=E_{\rm phys}^\perp\, e^{\hat t}\wedge e^{\hat\theta}.
\end{align}
Its nonzero coordinate components are
\begin{align}
F_{t\theta}&=E_{\rm phys}^\perp \sqrt{-g_{tt}g_{\theta\theta}},\nonumber\\
F_{\psi\theta}&=-E_{\rm phys}^\perp \frac{g_{t\psi}\sqrt{g_{\theta\theta}}}{\sqrt{-g_{tt}}},\nonumber\\
F_{\phi\theta}&=-E_{\rm phys}^\perp\frac{g_{t\phi}\sqrt{g_{\theta\theta}}}{\sqrt{-g_{tt}}}=\cos\theta\,F_{\psi\theta}.
\label{globalFTcomponents}
\end{align}
The additional spatial components in this equation are required by the nonzero time--space components of the metric. Inserting the full field strength, rather than retaining only its time--transverse component, is essential for obtaining the correct DBI reality condition.

Using these coordinate components in the DBI action, the determinant factorizes in terms of the physical field as $\det(g+2\pi\alpha' F^i)=\det(g)(1-\mathcal E_{\rm phys}^{i^2})$, where $i=\parallel,\perp$. Then, the critical electric fields in the local orthonormal frame become $E_{{\rm phys},c}^i=T_F$. The corresponding critical fields in the coordinate convention, expressed in terms of the physical boundary electric field, are therefore $E_{c,\mathrm{DBI}}^{\parallel,\mathrm{orth}}=E_{c,\mathrm{pot}}^\parallel$ and $E_{c,\mathrm{DBI}}^{\perp,\mathrm{orth}}=E_{c,\mathrm{pot}}^\perp$. Thus, the orthonormal-frame prescription reproduces the potential-analysis results in both directions, also in the presence of spatial--spatial cross-terms. The coordinate-field prescription, on the other hand, does not in general yield the correct critical field.

There is, however, a subtlety concerning the Bianchi identity for the longitudinal field. The two-form obtained in Eq.\,\eqref{globalFL} can be written as
\begin{align}
F=Q\,dt\wedge(d\psi+\cos\theta\,d\phi),
\qquad
Q=E_{\rm phys}^\parallel \sqrt{-g_{tt}h_{\psi\psi}},
\end{align}
where the metric components are evaluated at the position of the probe brane $z=z_0$. Thus, $Q$ is constant on the probe brane. Due to the angular dependence of the one-form $d\psi+\cos\theta\,d\phi$ associated with the longitudinal direction, this expression does not satisfy the Bianchi identity, $dF=0$. For the purpose of determining the critical electric field, however, this does not cause a problem. The DBI action is used here only to obtain the local reality condition, rather than to construct a global worldvolume gauge-field configuration. In a neighborhood of any chosen angular position $\theta_0$, we may therefore consider the closed two-form
\begin{align}
F_{\theta_0}=Q\,dt\wedge(d\psi+\cos\theta_0\,d\phi),
\end{align}
where $\theta_0$ is held fixed. At $\theta=\theta_0$, its coordinate components coincide with those of the field obtained from the orthonormal frame. Consequently, the DBI determinant and the resulting critical field at that position are unchanged. Furthermore, the critical field obtained in this way is independent of $\theta_0$, and hence gives the same result at any angular position. This local construction is therefore sufficient for the critical-field calculation considered here; a globally defined worldvolume gauge-field configuration would require a separate treatment.

\begin{figure}[t]
\centering
\includegraphics[width=0.48\textwidth]{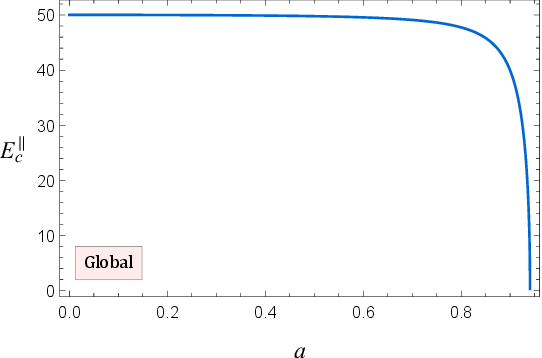}
\includegraphics[width=0.48\textwidth]{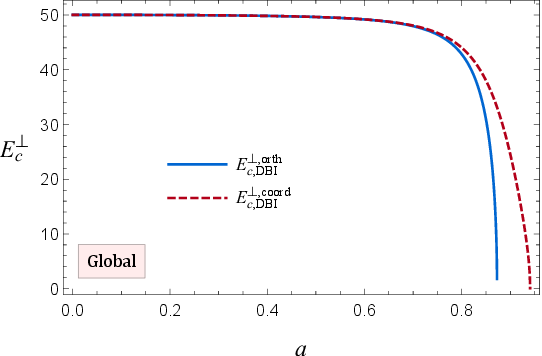}
\caption{\footnotesize Critical electric field as a function of the rotation
parameter $a$ in the global rotating geometry, for $T=1\,\mathrm{GeV}$ and $z_0=0.1\,\mathrm{GeV}^{-1}$. The left and right panels show the longitudinal and transverse directions, respectively. In the longitudinal panel, the potential-analysis result and the
orthonormal-frame DBI result coincide with the coordinate-field DBI result evaluated at the equator. In the transverse panel, the potential-analysis and orthonormal-frame results coincide, whereas the coordinate-field prescription gives a larger critical field for nonzero rotation.}
\label{Ecglobal}
\end{figure}

Figure \ref{Ecglobal} compares the critical fields as functions of the rotation parameter $a$ for $T=1\,\mathrm{GeV}$ and $z_0=0.1\,\mathrm{GeV}^{-1}$. In the longitudinal direction, the three results coincide at the equator, and the corresponding critical field
decreases monotonically with increasing rotation. It eventually vanishes when the black-hole horizon approaches the probe-brane position, $z_h\to z_0$. In the transverse direction, the
potential-analysis and orthonormal-frame results again coincide and decrease with $a$. In this case, the critical field vanishes when the zero of $g_{tt}$ reaches the probe brane, namely when $z_*\to z_0$. The coordinate-field result is larger than these two coincident curves for nonzero rotation. This result vanishes at a larger value of $a$ when $z_h \to z_0$. The difference increases with $a$, although it remains relatively small over much of the parameter range considered. All three transverse curves coincide in the nonrotating limit $a \to 0$.

For the chosen temperature, the system remains in the deconfined phase throughout the allowed range of $a$. Indeed, as can be seen in Eq.\,\eqref{Tcglobal}, the critical Hawking--Page temperature decreases monotonically with $a$ and satisfies $T_{\mathrm{HP}}(0) =\frac{3}{2\pi R}$. With $R=1\,\mathrm{GeV}^{-1}$, the chosen value $T=1\,\mathrm{GeV}$ is therefore above $T_c(a)$ for all $a\geq0$ in the physical range.

We now turn to the separation length and the total potential of the quark--antiquark pair. For the global rotating geometry, the functions entering the general expressions derived in
Sec.\,2 are
\begin{align}
A^{\rm global}(z)&=-g_{tt}g_{zz}=\frac{R^4}{z^4G(z)}
\left(1+\frac{R^2}{z^2}-\frac{2\mu z^2}{R^4}\right),
\label{globalA}
\end{align}
and functions $B_i^{\rm global}(z)$, written explicitly in Eqs.\,\eqref{globalBL} and \eqref{globalBT}. Substituting these functions into the general expressions for the separation length and the string contribution to the potential, obtained in Eqs.\,\eqref{generalseparation} and \eqref{generalQpotential}, gives the corresponding results for the longitudinal and transverse configurations. The total potential is then obtained by including the contribution of the external electric field, as in the general definition given in Eq.\,\eqref{generaltotalpotential}.

\begin{figure}[t]
\centering
\includegraphics[width=0.48\textwidth]{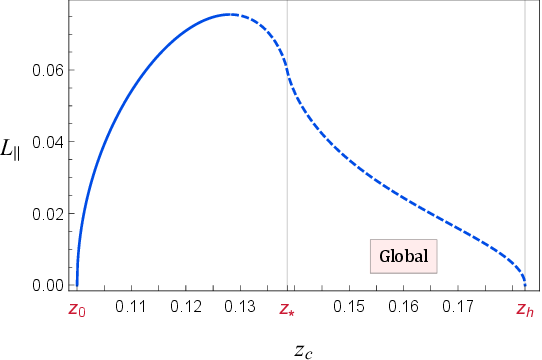}
\includegraphics[width=0.48\textwidth]{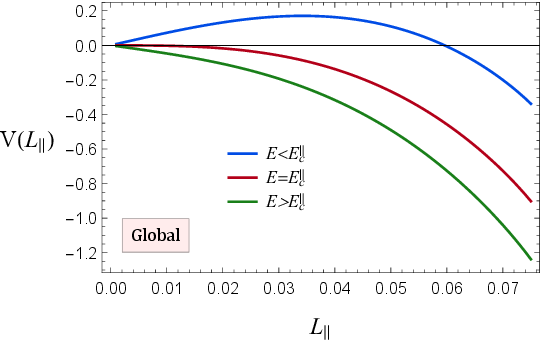}
\caption{\footnotesize Longitudinal quark--antiquark configuration in the
global rotating geometry for $T=1\,\mathrm{GeV}$, $z_0=0.1\,\mathrm{GeV}^{-1}$, and $a=0.8\,\mathrm{GeV}^{-1}$. The left panel shows the separation length as a function of
the turning point $z_c$, while the right panel shows the total potential for different values of the applied electric field. The barrier disappears at the common critical field obtained from the potential analysis, the orthonormal-frame DBI prescription and the coordinate-field DBI result at the equator.}
\label{LVLglobal}
\end{figure}

Figure~\ref{LVLglobal} shows the separation length and total potential for the longitudinal configuration, with $T=1\,\mathrm{GeV}$, $z_0=0.1\,\mathrm{GeV}^{-1}$, and $a=0.8\,\mathrm{GeV}^{-1}$. The left panel exhibits the two-branch structure characteristic of the deconfined phase. The second branch with the turning point farther from the boundary shows only a change in behavior around $z_*$, but the branch itself persists up to $z_h$. This is consistent with the fact that the longitudinal effective string tension does not vanish at $z_*$.

The right panel displays the total potential for several values of the applied electric field. As in the planar-limit geometry, the potential barrier decreases as the electric field is
increased and disappears at the critical field $E_{c,\mathrm{DBI}}^{\parallel,\mathrm{orth}}=E_{c,\mathrm{pot}}^\parallel$. Above this value, the total potential decreases monotonically, indicating the onset of the Schwinger pair-production process. Notice that $E_{c,\mathrm{DBI}}^{\parallel,\mathrm{coord}}$ does not correspond to the critical electric field in the longitudinal direction at arbitrary $\theta$.

\begin{figure}[t]
\centering
\includegraphics[width=0.48\textwidth]{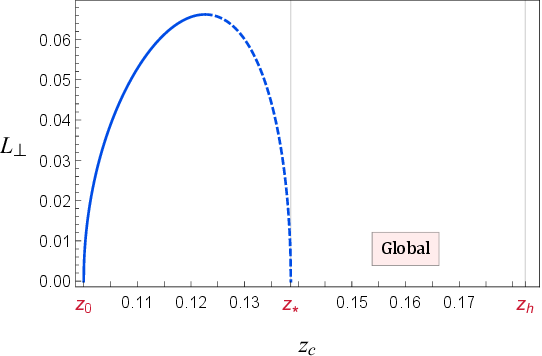}
\includegraphics[width=0.48\textwidth]{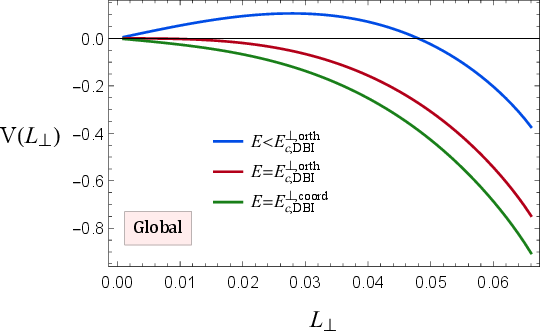}
\caption{\footnotesize Transverse quark--antiquark configuration in the
global rotating geometry for $T=1\,\mathrm{GeV}$, $z_0=0.1\,\mathrm{GeV}^{-1}$, and $a=0.8\,\mathrm{GeV}^{-1}$. The left panel shows the separation length as a function of
the turning point $z_c$, and the right panel shows the total potential for different values of the applied electric field, including $E_{c,\mathrm{DBI}}^{\perp,\mathrm{orth}}=E_{c,\mathrm{pot}}^\perp$, $E_{c,\mathrm{DBI}}^{\perp,\mathrm{coord}}>E_{c,\mathrm{pot}}^\perp$, and an electric field smaller than the transverse critical electric field.}
\label{LVTglobal}
\end{figure}

The corresponding transverse results are shown in Fig.\,\ref{LVTglobal}. The separation
length again exhibits a two-branch structure. However, unlike the longitudinal configuration, the second branch terminates at $z_*$, where the transverse effective string tension vanishes. Thus, the turning point of the transverse U-shaped string cannot extend beyond this surface. The total potential in the right panel shows the same qualitative behavior as in the longitudinal case: the potential barrier is reduced by the applied electric
field and disappears when the field reaches the critical value $E_{c,\mathrm{DBI}}^{\perp,\mathrm{orth}}=E_{c,\mathrm{pot}}^\perp$. At $E_{c,\mathrm{DBI}}^{\perp,\mathrm{coord}}$, the total potential is already decreasing and the pair production occurs without any obstacles. This shows that $E_{c,\mathrm{DBI}}^{\perp,\mathrm{coord}}$ does not corespond to the critical electric field in the transverse direction.

%**************************************************************************
\section{Discussion and final remarks}

In this work, we have formulated the DBI critical-field condition for stationary non-diagonal geometries by defining the applied electric field in the local orthonormal frame of a specified observer. We have shown that a field which is purely electric in this frame can acquire additional components when expressed in the coordinate basis. These components can include both additional temporal-spatial and spatial--spatial components, and must be retained when constructing the DBI matrix. Consequently, applying the coordinate-field prescription commonly used for diagonal geometries can lead to an incorrect critical field. This issue follows from the tensorial structure of the DBI action in a non-diagonal geometry and is not specific to a particular black-hole solution.

By transforming the complete field-strength tensor from the local orthonormal frame to the coordinate basis, we obtained the DBI reality condition for the geometries considered in this work. The critical field measured in the orthonormal frame takes the universal value
\begin{align}
E_{\mathrm{proper},c}=T_F,
\end{align}
where $T_F$ is the fundamental string tension. The dependence on the background geometry enters through the relation between this locally measured field and its coordinate components. We also derived the critical field independently from the holographic potential analysis. In all general cases considered, the disappearance of the potential barrier occurs at the same field strength as the one obtained from the orthonormal-frame DBI determinant. We demonstrated this agreement in the planar limit of the rotating Myers--Perry geometry and in the global Myers--Perry--AdS geometry. The latter provides a more general example in which both temporal--spatial and spatial--spatial cross-terms are present. In contrast, the coordinate-field prescription does not in general reproduce this agreement, since the coordinate components used in that prescription need not represent the intended physical electric field.

We have also encountered a subtlety related to the Bianchi identity in the global Myers--Perry geometry. The longitudinal field obtained from the local orthonormal frame does not, in its original form, define a closed two-form over the full angular space. This is due to the angular dependence of the one-form associated with the longitudinal direction. Since our aim is to determine the local critical electric field, a closed extension can instead be considered in a neighborhood of any chosen angular position. The resulting critical field is independent of this position and therefore gives the same value throughout the angular space. This issue is not restricted to non-diagonal geometries: even for a diagonal metric, the Bianchi identity may fail for a field defined as purely electric in a local orthonormal frame if the metric components entering that field depend on spatial coordinates transverse to its direction. In such cases, the spatial dependence of the orthonormal-frame components can prevent the corresponding two-form from being globally closed. A global realization of the field would then require additional considerations, which are not needed for the local critical-field analysis pursued here.

The prescription developed in this work applies to stationary geometries for which a regular local orthonormal frame can be defined with respect to a specified timelike observer. For the static frame considered here, this requires the stationary time direction to be timelike in the region of interest and the induced spatial metric in the observer's rest frame to be positive definite. When these conditions are not satisfied, an alternative timelike observer must be specified, and the electric field must be defined in that observer's orthonormal frame. Our analysis has been restricted to a single applied electric field in a specified spatial direction, without additional independent components of the worldvolume field strength. Including background magnetic fields or other electric components may modify the DBI reality condition and requires treating the full worldvolume field-strength tensor. Extending the present analysis to such configurations, together with the construction of globally defined worldvolume gauge fields, provides a natural direction for future work.
%*********************************************************************************************************************
%\section*{Acknowledgement}
%\appendix
%\section*{Appendix A} \label{Calculation}
%\setcounter{equation}{0}
%\renewcommand{\theequation}{\Alph{section}.\arabic{equation}}

 \end{document}